\documentclass[aps,reprint,superscriptaddress]{revtex4-1}
\usepackage{xcolor}
\usepackage{xfrac}
\usepackage{amsmath}
\usepackage{graphicx}
\usepackage{upgreek}
\usepackage[yyyymmdd,hhmmss]{datetime}
\usepackage{lmodern}
\usepackage[T1]{fontenc}
\usepackage{soul}

\newcommand{\vect}[1]{\boldsymbol{\mathbf{#1}}}

\begin{document}

\title{Designing Nanomagnet Arrays for Topological Nanowires in Si}

\author{L. N. Maurer}
\email{lmaurer@sandia.gov}
\affiliation{Sandia National Laboratories, Albuquerque, New Mexico 87185, USA}

\author{J. K. Gamble}
\altaffiliation{Currently with Microsoft Research, Redmond, WA, USA}
\affiliation{Sandia National Laboratories, Albuquerque, New Mexico 87185, USA}

\author{L. Tracy}
\affiliation{Sandia National Laboratories, Albuquerque, New Mexico 87185, USA}

\author{S. Eley}
\affiliation{Los Alamos National Lab, Los Alamos, New Mexico 87545, USA}

\author{T. M. Lu}
\email{tlu@sandia.gov}
\affiliation{Sandia National Laboratories, Albuquerque, New Mexico 87185, USA}

\begin{abstract}

Recent interest in topological quantum computing has driven research into topological nanowires, one-dimensional quantum wires that support topological modes including Majorana fermions. Most topological nanowire designs rely on materials with strong spin-orbit coupling, such as InAs or InSb, used in combination with superconductors. It would be advantageous to fabricate topological nanowires using Si owing to its mature technology. However, the intrinsic spin-orbit coupling in Si is weak.  One approach that could circumvent this material deficiency is to rotate the electron spins using nanomagnets. Here, we perform detailed simulations of realistic Si/SiGe systems with an artificial spin-orbit gap induced by a nanomagnet array. Most of our results are also generalizable to other nanomagnet-based topological nanowire designs. By studying several concrete examples, we gain insight into the effects of nanomagnet arrays, leading to design rules and guidelines. Finally, we present an experimentally realizable design using magnets with a single polarization.

\end{abstract}

\maketitle

\section{Introduction}

There is an ongoing search for systems that support Majorana fermions because of their potential in fault-tolerant topological quantum computing \cite{Sau_Physics_17}. A particularly promising class of systems are topological nanowires: heterostructures made from conventional s-wave superconductors combined with nanowires made from a material with strong spin-orbit coupling, such as InAs or InSb, placed in an external magnetic field.  The external field, perpendicular to the spin-orbit field, mixes the two spin subbands near $k=0$ and creates a spin-orbit Zeeman gap.  Inside this gap, the spin degeneracy is lifted, leading to Majorana modes at the ends of the nanowires when superconductivity is induced by the superconducting proximity effect \cite{Sarma_PRL_10, Oppen_PRL_10}. Many experimental studies of topological nanowires have shown signatures of topological modes \cite{Kouwenhoven_Sci_12, Shtrikman_NPhys_12, Furdyna_NPhys_12, Xu_NLett_12, Li_PRL_13, Marcus_PRB_13, Kouwenhoven_NCom_17}.

However, it would be convenient if the nanowires could be made from materials without strong spin-orbit coupling, such as Si, where industrial applications have developed much more mature fabrication capabilities than with InAs or InSb. It is important to realize that materials with intrinsic spin-orbit coupling are not required for the formation of topologically non-trivial modes. The utility of the strong spin-orbit coupling is to offset the E-k dispersion from $k=0$ so that a spin-orbit gap opens when an external magnetic field is applied \cite{Sarma_PRL_10, Oppen_PRL_10}. However, a similar gap can be opened in other ways. In particular, there have been several designs proposed that use nanomagnet arrays to engineer an artificial spin-orbit gap in materials without strong spin-orbit coupling \cite{Loss_PRL_12, Flensberg_PRB_12, Ojanen_PRB_13, Cuoco_PRB_17}.

\begin{figure}
 \centering
 \includegraphics[width=0.99\columnwidth]{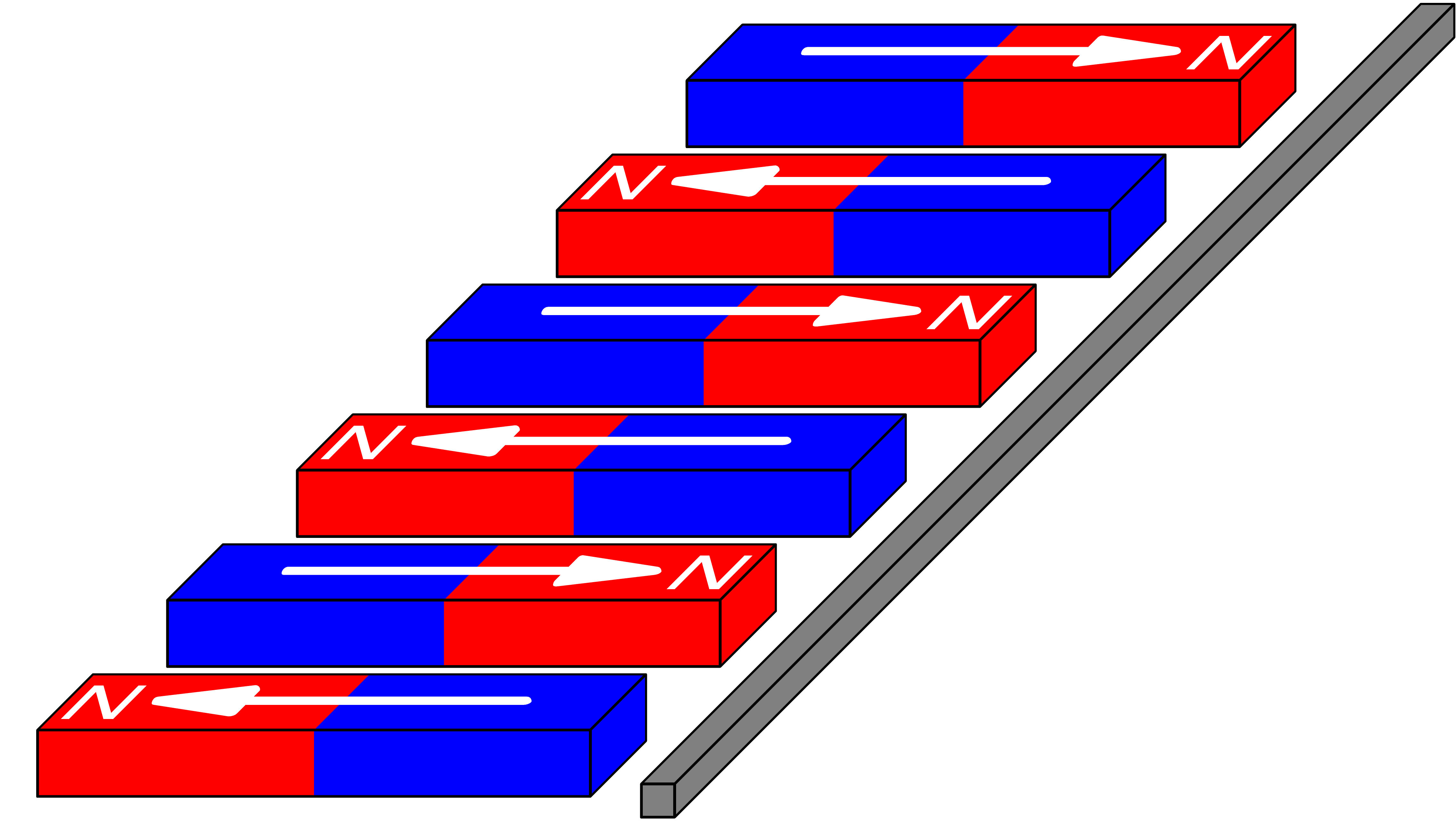}
 \caption{Basic layout of a system with an artificial spin-orbit gap. A 1D quantum wire with weak intrinsic spin-orbit coupling (grey) is next to an array of nanomagnets with alternating polarizations. The magnets produce a rotating magnetic field in the wire that creates a spin-orbit gap in the band structure \cite{Loss_PRB_10, Flensberg_PRB_12}.} \label{fig:rough_device}
\end{figure}

One design for a system with an artificial spin-orbit gap consists of a one-dimensional (1D) quantum wire next to a nanomagnet array \cite{Loss_PRL_12, Flensberg_PRB_12} (Fig.~\ref{fig:rough_device}). Designs have been proposed where the magnets alternate polarization \cite{Loss_PRL_12, Flensberg_PRB_12} or are all polarized in the same direction \cite{Flensberg_PRB_12}. Klinovaja et al. assumed that the nanomagnets would produce a perfectly sinusoidal magnetic field in the wire \cite{Loss_PRL_12}, while Kjaergaard et al. calculated the actual magnetic field for two nanomagnet array geometries \cite{Flensberg_PRB_12} but did not report the band structures.

In this paper, we perform detailed calculations of the magnetic fields and band structures resulting from realistic nanomagnet array geometries. This allows us to see how design realities introduce imperfections into the band structure and gives us insight into how to mitigate the imperfections. In particular, we illustrate a promising design where all magnets have the same polarization. We also give simple design advice for how to eliminate unwanted gaps in the band structure (Sec.~\ref{sec:gaps}), examine the effects of using magnets of different strengths in one device (Sec.~\ref{sec:strengths}), and study different designs where all magnets have the same polarization (Sec.~\ref{sec:unipolarized}).








\section{System and Methods}

\begin{figure}
 \centering
 \includegraphics[width=0.85\columnwidth]{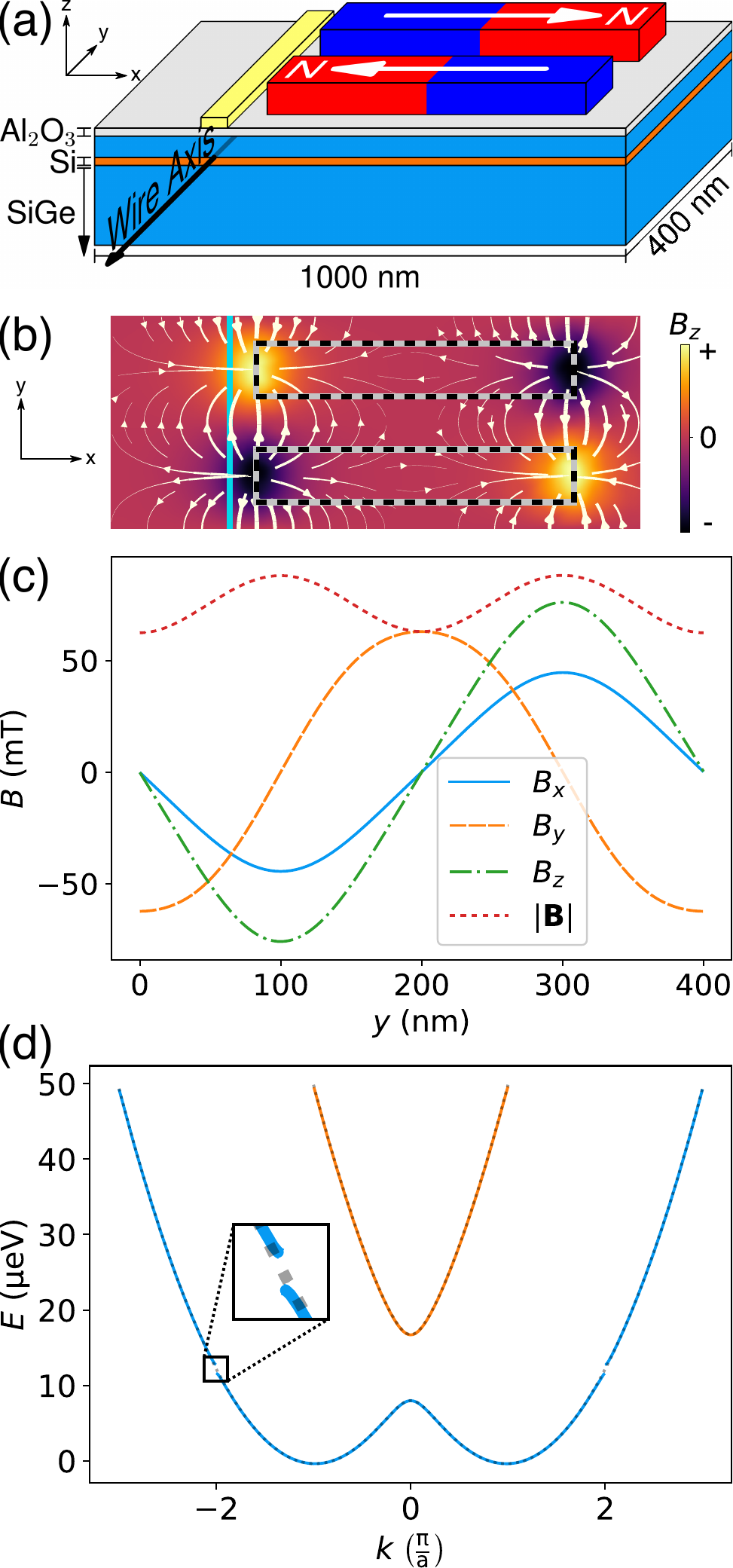}
 \caption{(a) A unit cell of an example device design. The nanomagnets are 600 nm long, 100 nm wide, have alternating polarization, and are placed on a stack consisting of a 15-nm-thick $\text{Al}_2\text{0}_3$ layer (grey), 40-nm-thick SiGe barrier (blue), 15-nm-thick Si well (orange), and a thick SiGe buffer (blue). An accumulation gate (yellow) generates a 1D channel (wire) in the Si layer directly below the gate. The axis of this wire is indicated. An actual device would have many of these unit cells tiled along the wire axis. (b) The magnetic field in the plane of the quantum well. Arrows indicate in-plane field. Colors represent out-of-plane field. The wire axis is in blue. The magnets above the well are outlined in dashed black and grey. (c) The magnetic field along the wire axis with 2.2 T magnets. (d) The resulting band structure after the unitary transformation  (Sec.~\ref{sec:U}). The dotted, light-grey curve shows the band structure resulting from a perfectly sinusoidal magnetic field [Eq.~(\ref{eq:E_SOC})] with $B$ equal to the average $\left|\vect{B}\left(y\right)\right|$. The light-grey curve is an excellent match except for two small gaps (insert), which are explained in Sec.~\ref{sec:gaps}.} \label{fig:device}
\end{figure}

\subsection{Physical System} \label{sec:system}

We consider systems like the one shown in Fig.~\ref{fig:device}, where electrons in a Si channel (wire) travel through an engineered, rotating magnetic field. Here, we use undoped Si/SiGe as the model material system, as it allows for high electron mobilities \cite{Lu_APL_09, Lu_Adv_15,Petta_PRB_15} and, in turn, mean free paths longer than the length of a magnetic field rotation. However, our results are general and can be applied to other nanomagnet/1D wire systems.

In our design, an accumulation gate creates a quasi-1D electron channel (wire) in a Si quantum well, and a nanomagnet array generates a rotating magnetic field. An actual device would have many of the unit cells shown in Fig.~\ref{fig:device}(a) tiled along the wire axis. Because the size of the electron channel is tunable with the accumulation gate, we can ensure that there are not too many transverse modes in the channel. In the limit of an extremely narrow gate, the number of modes is limited by the thickness of the SiGe cap due to smearing of the electrostatic potential.  At 40 nm, single-mode occupancy is achievable. 

We present devices where the magnets have alternating polarizations and devices where all the magnets have a single polarization. Alternating-polarization devices could be fabricated with a two-step process using two different magnetic materials with different coercivities, such as Co and SmCo. First, all the magnets would be polarized in one direction by a magnetic field larger than the coercivities of the two materials, then the magnets with the weaker coercivity would be polarized in the opposite direction by a magnetic field between the two coercivities. In our simulations, we assume that the remanent magnetization for Co and SmCo magnets are 2.2 T and 1.0 T, respectively. (These numbers were obtained from magnetic moment measurements of Co and SmCo thin films at 4 K.) Unless otherwise noted, all nanomagnets in our designs are 60 nm thick.

\subsection{Theoretical Model} \label{sec:model}

We only investigate the spin-orbit gap caused by nanomagnet arrays; in this work we do not consider the superconductors necessary for topological nanowires. The Hamiltonian for our system is then

\begin{equation} \label{eq:H}
  H = \frac{\hat{p}^2_y}{2 m} + \frac{g \mu_B}{2} \vect{B}\left(y\right) \cdot \vect{\sigma},
\end{equation}

\noindent
where $y$ is the position along the axis of the wire, $m = 0.19 \; m_0$ is the effective mass, $g$ is the Land\'e g-factor, $\mu_B$ is the Bohr magneton, $\vect{B}$ is the magnetic field at the wire, and $\vect{\sigma}$ is the array of Pauli matrices.

In order for a design to create a useful artificial spin-orbit gap, the magnetic field, when viewed in some plane, must rotate a full $2 \pi$ over one unit cell of the structure. I.e., as we move down the wire, the direction of the field must cycle $\rightarrow, \downarrow, \leftarrow, \uparrow$; alternating $\uparrow, \downarrow$ or $\leftarrow, \rightarrow$ will not suffice (Sec.~\ref{sec:gaps}).

To model the system, we start by calculating the magnetic field using \textsc{comsol multiphysics} \cite{COMSOL}. Our simulation has periodic boundary conditions along the wire axis and infinite-element domains \cite{Bettess_book} at the other edges.

We then use the magnetic field to calculate the band structure using the finite difference method described in Sec.~\ref{sec:FD}. In Sec.~\ref{sec:U}, we discuss the unitary transformation used to compare our system to one with spin-orbit coupling in an external magnetic field.

\subsection{Finite Difference Solution for Band Structure} \label{sec:FD}

We solve for the dispersion relation $E_n\left(k\right)$, where $n$ is the band index, by discretizing the Hamiltonian [Eq.~(\ref{eq:H})] into a finite difference equation and applying Bloch boundary conditions:

\begin{align} \label{eq:H_disc}
  \begin{split}
 		H\left(k\right) = \frac{1}{2m\Delta^2}
		\begin{bmatrix}
			-2         & 1      &        & e^{ika} \\
			1          & \ddots & \ddots &         \\
			           & \ddots & \ddots &  1      \\
			e^{-ika}   &        & 1      & -2      \\
		\end{bmatrix} &\otimes I \\
	+ \frac{g \mu_B}{2} \sum_{i=x,y,z}
		\begin{bmatrix}
			B_{i,0} &        &         \\
			        & \ddots &         \\
			        &        & B_{i,N} \\
		\end{bmatrix} &\otimes \sigma_i,
	\end{split}
\end{align}

\noindent
where we have discretized the system into $N$ elements of width $\Delta = \frac{a}{N}$, $a$ is the length of one unit cell, $B_{i,m}$ is the `i'th component of the magnetic field at position $m\Delta$ for $m \in  \left[0,N\right]$, and $I$ is the 2x2 identity matrix. Both matrices in brackets are of size NxN with zeros everywhere except indicated. The $e^{\pm ika}$ terms implement the Bloch boundary conditions that enforce $\psi\left(a\right) = e^{ika}\psi\left(0\right)$, which is required by Bloch's theorem. The dispersion relation $E_n\left(k\right)$ is given by the eigenvalues of $H\left(k\right)$ [Eq.~(\ref{eq:H_disc})].

\subsection{Unitary Transformation for a Perfectly Sinusoidal Magnetic Field} \label{sec:U}

Both Refs. \cite{Loss_PRB_10, Flensberg_PRB_12} use a position-dependent unitary transformation $U\left(y\right)$ to convert the Hamiltonian in Eq.~(\ref{eq:H}) into $\mathcal{H} = U^{\dagger}HU$, which for an appropriate magnetic field, has the form expected from a system with Rashba-type spin-orbit coupling in an external magnetic field. The appendix to this paper reviews the transformation used in \cite{Flensberg_PRB_12}.

Here we consider the unitary transformation resulting from

\begin{equation} \label{eq:sinusoidal_B}
  \vect{B}\left(y\right) = B \, \sin\left(\frac{2 \pi y}{a}\right) \hat{\vect{x}} + B \, \cos\left(\frac{2 \pi y}{a}\right) \hat{\vect{y}},
\end{equation}

\noindent
which we will call a \emph{perfectly sinusoidal} magnetic field. Plugging Eq.~(\ref{eq:sinusoidal_B}) into the general unitary transform [Eqs.~(\ref{eq:U})-(\ref{eq:variable_theta})] results in

\begin{equation} \label{eq:sinusoidal_U}
  U\left(y\right) = e^{i \frac{\pi y}{a} \sigma_z } = \cos{\left(\frac{\pi y}{a}\right)} I + i \, \sin{\left(\frac{\pi y}{a}\right)} \sigma_z.
\end{equation}

\begin{figure}
 \centering
 \includegraphics[width=0.99\columnwidth]{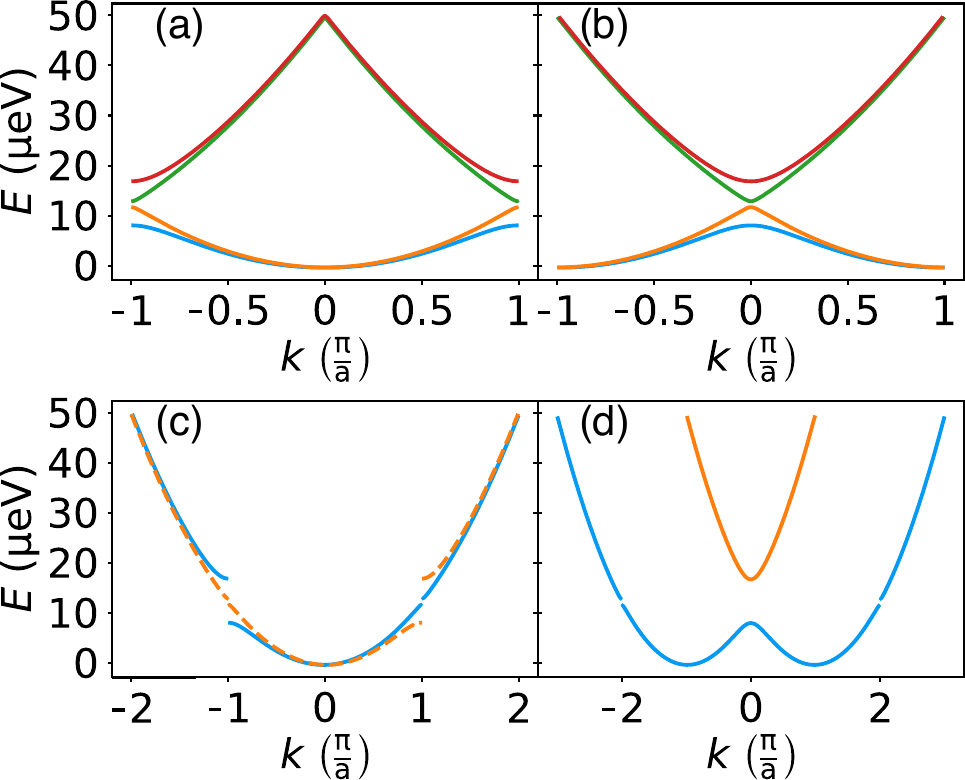}
 \caption{Band structure from the device in Fig.~\ref{fig:device} shown in four ways: without (left) and with (right) the unitary transform [Eq.~(\ref{eq:sinusoidal_U})] applied. Reduced-zone scheme (top) and extended-zone scheme (bottom) \cite{Ashcroft_Mermin}. The orange curve in the lower left (reduced-zone scheme without transformation) is dashed to make the overlap with the blue curve clear.} \label{fig:transform_and_zone}
\end{figure}

Even when the magnetic field is not perfectly sinusoidal, we are free to use the unitary transformation in Eq.~(\ref{eq:sinusoidal_U}), which converts the band structure resulting from our Hamiltonian [Eq.~(\ref{eq:H})] to a band structure similar to that in a system with strong spin-orbit coupling in an external magnetic field (Fig.~\ref{fig:transform_and_zone}). We use the unitary transformation in Eq. (\ref{eq:sinusoidal_U}) unless otherwise noted. For computational purposes, the unitary transformation can be discretized and applied to the discretized Hamiltonian in Eq.~(\ref{eq:H_disc}).

\section{Results} \label{sec:results}

By calculating band structures from a variety of example device designs, we have developed insight into the difficulties of realistic designs and design principles to tackle the challenges. In this section, we provide results for example designs that illustrate these challenges and workarounds. Our results show that realistic designs should feasibly produce a measurable artificial spin-orbit gap.

Because we are trying to reproduce the band structure given by a system like an InAs or InSb nanowire, we look at two factors to evaluate a device: the size of the spin-orbit gap, and how close the band structure is to the band structure of an ideal system with spin-orbit coupling in an external magnetic field. However, we note that nanomagnet arrays are very flexible and can foster topological states that are much different from the topological states found in InAs or InSb nanowires (e.g., \cite{Ojanen_PRB_13}), so our approach to evaluating designs is relatively conservative.

A system with a perfectly sinusoidal magnetic field reproduces the ideal system and results in the band structure \cite{Flensberg_PRB_12}

\begin{equation} \label{eq:E_SOC}
  E = \frac{\hbar^2 k^2}{2 m} \pm \sqrt{\left(\frac{g \mu_B B}{2}\right)^2 + \left(\frac{\hbar^2 k}{2 m} k_0 \right)^2} + \frac{\hbar^2}{8m} k_0^2,
\end{equation}

\noindent
where $k_0 = \frac{2 \pi}{a}$. So, in most of the band structure plots, we will overlay the curves from Eq.~(\ref{eq:E_SOC}) on the calculated band structure, which we generally plot in the extended-zone scheme with the unitary transformation [Fig.~\ref{fig:transform_and_zone}(d)]. To make the comparison between the computed band structure and Eq.~(\ref{eq:E_SOC}) useful, we choose $B$ in Eq.~(\ref{eq:E_SOC}) to be the average magnetic field magnitude at the wire:

\begin{equation}
	B = \frac{1}{a}\int_{\substack{\text{unit} \\ \text{cell}}} \left|\vect{B}\left(y\right)\right| dy
\end{equation}

Take the example of the device shown in Fig.~\ref{fig:device}(a). The calculated band structure is shown in Fig.~\ref{fig:device}(d). Eq.~(\ref{eq:E_SOC}) is overlaid as a dotted grey curve. The calculated and ideal band structure are virtually identical except for two small gaps (magnified in the inset) that open in the blue curve. The main spin-orbit gap has an energy of 8.7 {\textmu}eV, which corresponds to 101 mK. While small, that gap should be measurable in a dilution refrigerator.

In Sec.~\ref{sec:gaps}, we study the origin of the gaps seen in Fig.~\ref{fig:device}(d). In Sec.~\ref{sec:strengths}, we investigate the effects from having a nonzero average magnetic field at the wire. In Sec.~\ref{sec:unipolarized}, we consider designs where all the magnets are polarized in the same direction, which would likely be easier to fabricate than alternating-polarization designs. Finally, in~\ref{sec:spin} we see how the spin of an electron rotates as it moves through the wire.

\subsection{Controlling Gaps and Design Advice} \label{sec:gaps}

There are two important gaps in the band structure shown in Fig.~\ref{fig:device}(d): the main spin-orbit gap at $k=0$ (the \emph{good} gap) and smaller gaps around the same energy (the \emph{bad} gaps). The good gap defines the energy window where the system is spinless, and the bad gap reduces the size of it.  While the presence of the bad gap does not prevent induced superconductivity, it is undesirable for practical purposes.  A bad gap smaller than $k_BT$, where $k_B$ is the Boltzmann constant and $T$ is the measurement temperature, can likely be ignored in experiments.  We consider the conditions needed to eliminate the bad gaps and give design advice at the end of this subsection.

In particular, we will show that, if the magnetic field is in a plane, the bad gap will be zero if the two in-plane components are identical in magnitude and 90 degrees out of phase,  i.e., a magnetic field rotating in the plane. This follows from the fact that both gaps are controlled by the first Fourier coefficient of the magnetic field $\vect{b}_{K_1}$, where the magnetic field can be expanded as a Fourier series

\begin{equation}\label{eq:B_expansion}
  \vect{B}\left(y\right) = \sum_n \vect{b}_{K_n} e^{i K_n y},
\end{equation}

\noindent
with $K_n = \frac{n \pi}{a}$ a reciprocal lattice vector.

When the raw band structure is plotted in the reduced-zone scheme, both gaps appear at the edge of the first Brillouin zone [Fig.~\ref{fig:transform_and_zone}(a)]. So, these gaps can be analyzed the same way as energy levels near a single Bragg plane (Ref. \cite{Ashcroft_Mermin} Ch. 9, although the method must be extended to account for spin).

The potential energy term $U_E\left(y\right) = \frac{g \mu_B}{2} \vect{B}\left(y\right) \cdot \vect{\sigma}$ in Eq.~(\ref{eq:H}) can be expanded as

\begin{equation} \label{eq:U_expansion}
  U_E\left(y\right) =  \sum_n u_{k_n} e^{i K_n y}
\end{equation}

\noindent
where the 2x2 matrix 

\begin{equation} \label{eq:u_k}
  u_{K_n} = \frac{g \mu_B}{2}  \vect{b}_{K_n} \cdot \vect{\sigma}.
\end{equation}

Using degenerate perturbation theory, the four energies $E_{0\dots3}$ of the states at the edge of the Brillouin zone are solutions of [Ref. \cite{Ashcroft_Mermin} Eq. (9.24)]:

\begin{align} 
  \begin{split}
    0 & =
    \begin{vmatrix}
      \left(E_n - E_{K_1}^0\right) I & -u_{K_1} \\
      -u_{K_1}^{\dagger} & \left(E_n - E_{K_1}^0\right)I 
    \end{vmatrix}, \\
    0 & = \left| \left(E_n - E_{K_1}^0\right)^2 I - {u_{K_1}}^{\dagger} u_{K_1}\right|,
  \end{split}
\end{align}

\noindent
where $E_{K_1}^0 = \frac{\hbar^2}{2 m} \left(\frac{\pi}{a}\right)^2$ is the energy of a free particle with wavenumber $K_1$. The size of the two gaps are $G_0 = E_2 - E_1$ and $G_1 = E_3 - E_0$ when the $E_n$ are in increasing order. Equivalently, the gaps are given by

\begin{equation}
  G_i = 2 \sqrt{\lambda_i},
\end{equation}

\noindent
where $\lambda_i$ are the eigenvalues of the matrix ${u_{K_1}}^{\dagger} u_{K_1}$.

Hence, the size of the gaps is determined by the first Fourier coefficient of the magnetic field $\vect{b}_{K_1}$, which is contained in $u_{K_1}$. The larger of the gaps ($G_1$) is the good gap, and the smaller ($G_0$) is the bad gap.

Consider the situation where the magnetic field lies in a plane (chosen to be the x-y plane). If the x and y components of $\vect{b}_{K_1}$ have the same magnitude but are out of phase by an angle $\phi$, then we can write $\vect{b}_{K_1} = c \hat{\vect{x}} + c e^{i\phi} \hat{\vect{y}}$, where $c$ is some constant. Combined with Eq.~(\ref{eq:u_k}), we have

\begin{align}
  \begin{split}
               u_{K_1} & = \frac{g \mu_B}{2} \left(c \hat{\vect{x}} + c e^{i\phi} \hat{\vect{y}} \right) \cdot \vect{\sigma},\\
                       & = \frac{g \mu_B}{2} c \left(\sigma_x + e^{i\phi}\sigma_y\right), \\
                       & = \frac{g \mu_B}{2} c
                         \begin{bmatrix}
                           0 & 1-ie^{i\phi} \\
                           1+ie^{i\phi} & 0 
  					            \end{bmatrix},
  \end{split}
\end{align}

\noindent
and the two gaps are then

\begin{equation} \label{eq:gaps}
  G_{0,1} = g \mu_B c \sqrt{2 \pm 2\sin{\phi}}.
\end{equation}

So, if $\phi = \pm \sfrac{\pi}{2}$ (i.e., the two components are 90 degrees out of phase), then there is one gap of size $2 g \mu_B c$ and the other gap of size zero. In other words, the bad gap has been eliminated. Moving $\phi$ away from $\pm \sfrac{\pi}{2}$ increases the size of the bad gap while simultaneously decreasing the size of the good gap. This explains why a field that simply alternates $\uparrow, \downarrow$ will not work for our purposes: an alternating field has all its components in phase ($\phi = 0$), so both gaps will be the same size, and the device will not be useful.

This analysis gives us a recipe for maximizing the good gap and minimizing or eliminating the bad gap when the magnetic field is in a plane:

\begin{itemize}
  \item Equalize the magnitude of the x and y components of the first Fourier coefficient of the magnetic field $\vect{b}_{K_1}$.
  \item Make the x and y components of $\vect{b}_{K_1}$ be 90 degrees out of phase.
  \item Maximize $\left|\vect{b}_{K_1}\right|$.
\end{itemize}

A perfectly sinusoidal magnetic field [Eq.~(\ref{eq:sinusoidal_B})]  meets those conditions, but so do many other magnetic fields.

\subsection{Magnets of Different Strengths and the Average Magnetic Field} \label{sec:strengths}

\begin{figure}
 \centering
 \includegraphics[width=0.99\columnwidth]{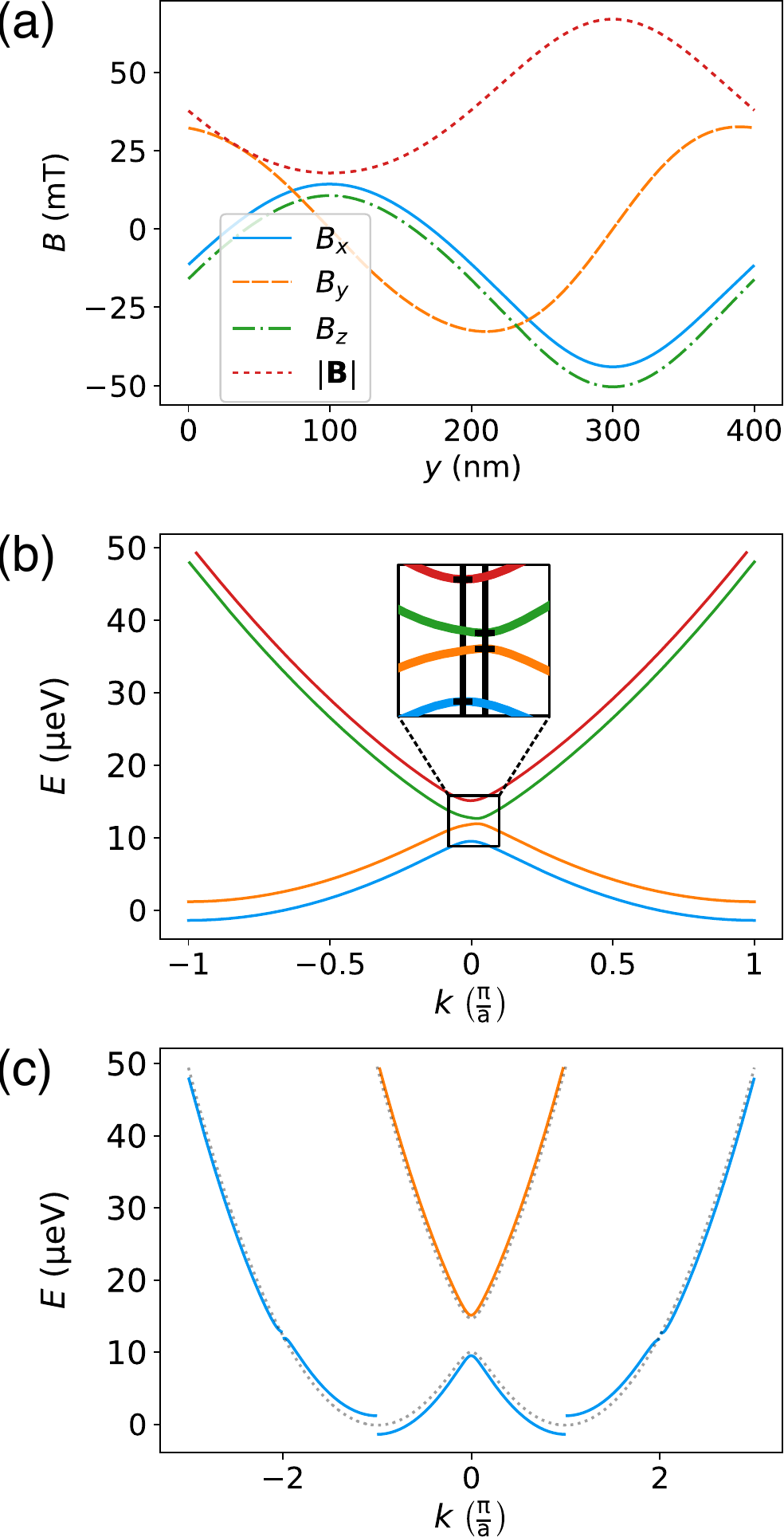}
 \caption{Results from the device in Fig.~\ref{fig:device}(a) where the two magnets of opposite polarization have different strengths of 2.2 and 1.0 T (appropriate for Co and SmCo), respectively. (a) The magnetic field along the wire. (b) The transformed band structure in the reduced-zone scheme. Compared with Fig.~\ref{fig:transform_and_zone}(b), the blue and orange curves have moved apart due to the average magnetic field (and the same for the red and green curves). The asymmetry in the magnetic field also caused the band maxima and minima to shift slightly left or right. Inset shows that the k values of band extrema (marked by vertical lines) have shifted and do not occur at the same k value. (c) The band structure in the extended-zone scheme after the unitary transformation shows that the nonzero average magnetic field creates new gaps at $k = \pm \sfrac{\pi}{a}$. The dotted, light-grey curve shows the band structure resulting from a perfectly sinusoidal magnetic field [Eq.~(\ref{eq:E_SOC})] with $B$ equal to the average $\left|\vect{B}\left(y\right)\right|$.} \label{fig:different_strengths}
\end{figure}

We define the \emph{average magnetic field}

\begin{equation}
	\bar{\vect{B}} = \frac{1}{a}\int_{\substack{\text{unit} \\ \text{cell}}} \vect{B}\left(y\right) dy,
\end{equation}

\noindent
to be the average value of the magnetic field at the wire. So far, we have only looked at designs where the average magnetic field is zero. Real devices are likely to have a nonzero average magnetic field. For example, alternating-polarization devices may be made with two different magnetic materials with different remanent magnetizations (Sec.~\ref{sec:model}), which can easily produce magnets of different strengths and $\bar{\vect{B}} \neq 0$.

When the average magnetic field is zero, the lowest two bands are degenerate at $k=0$ ($k=\sfrac{\pi}{a}$) in the untransformed (transformed) reduced-zone scheme (top of Fig.~\ref{fig:transform_and_zone}). This is no longer the case when there is a nonzero average magnetic field. So long as the gap due to the average magnetic field $\approx g \mu_B \left|\bar{\vect{B}}\right|$ is small relative to the spin-orbit gap, the most noticeable effect of a nonzero average magnetic field is that some bands in the reduced-zone scheme will be shifted up in energy while others will be shifted down. New and larger gaps result. In particular, the blue and orange curves in the repeated-zone schemes shown at the top of Fig.~\ref{fig:transform_and_zone} will shift apart [compare with Fig.~\ref{fig:different_strengths}(b)]. When transformed and put into the extended-zone scheme, the effect is opening new gaps at $k=\pm \sfrac{\pi}{a}$.

For larger average magnetic fields, the bands separate so much that they no longer interact, and analysis of gaps in Sec.~\ref{sec:gaps}, which assumes that the bands are degenerate at the edge of the Brillouin zone, becomes inaccurate. The spin-orbit gap is also washed out. (An example is shown in the next section.)

Another side effect of magnets of different strengths is that $\left|\vect{B}\left(y\right)\right|$ is often asymmetric, which can shift the band structure in $k$. Take the example shown in Fig.~\ref{fig:different_strengths}. $\left|\vect{B}\left(y\right)\right|$ [the red curve in Fig.~\ref{fig:different_strengths}(a)] is clearly asymmetric: the magnetic field components have different magnitudes and are not exactly 90 degrees out of phase. In Fig.~\ref{fig:different_strengths}(b), the extrema of the red and blue curves are at different k values than the extrema of the green and orange curves (see inset). The shifts in Fig.~\ref{fig:different_strengths}(b) are relatively small, but the shifts can be much larger if $\left|\vect{B}\left(y\right)\right|$ is very asymmetric.

In any case, the main spin-orbit gap remains so long as the average magnetic field is not too large, so the new gap and any shifts may not matter for topological nanowire applications. If the new gap is a problem, it is straightforward to eliminate in systems with magnets of two different polarizations and strengths: simply make the weaker magnet larger (or the stronger magnet smaller) until the average magnetic field is zero. This will often also make $\left|\vect{B}\left(y\right)\right|$ more symmetric.

\subsection{Designs with Only One Magnet Polarization} \label{sec:unipolarized}

\begin{figure}
 \centering
 \includegraphics[width=0.90\columnwidth]{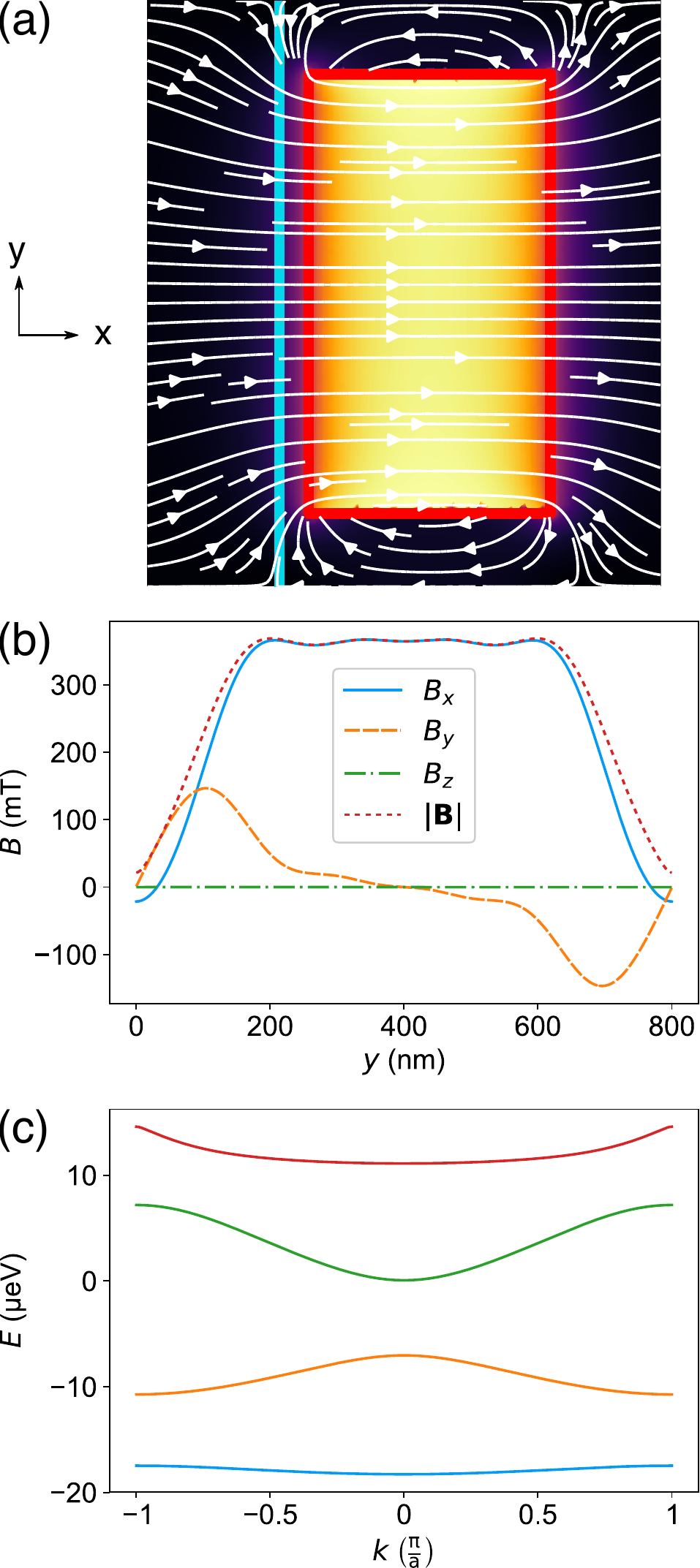}
 \caption{Device modeled after the single-polarization device in Ref. \cite{Flensberg_PRB_12}: an array of 600-nm-wide, 330-nm-long nanomagnets spaced 200 nm apart. The magnet is in plane with the quantum well. Ref. \cite{Flensberg_PRB_12} did not state the magnet thickness or remanent magnetization, so we have chosen 60 nm and 2 T, respectively. (a) Magnetic field through the plane of the quantum well. The magnet is outlined in red. A wire located 40 nm from the magnet is shown in blue. Other colors indicate the magnitude of the magnetic field (light high, dark low). (b) The magnetic field along the wire. It  does rotate along the wire axis. However, the magnitudes of the x and y components are very different, and the average magnetic field is far from zero. (c) The resulting band structure in the reduced zone scheme without transformation. There is no identifiable spin-orbit gap.} \label{fig:PRB_device}
\end{figure}

\begin{figure}
 \centering
 \includegraphics[width=0.85\columnwidth]{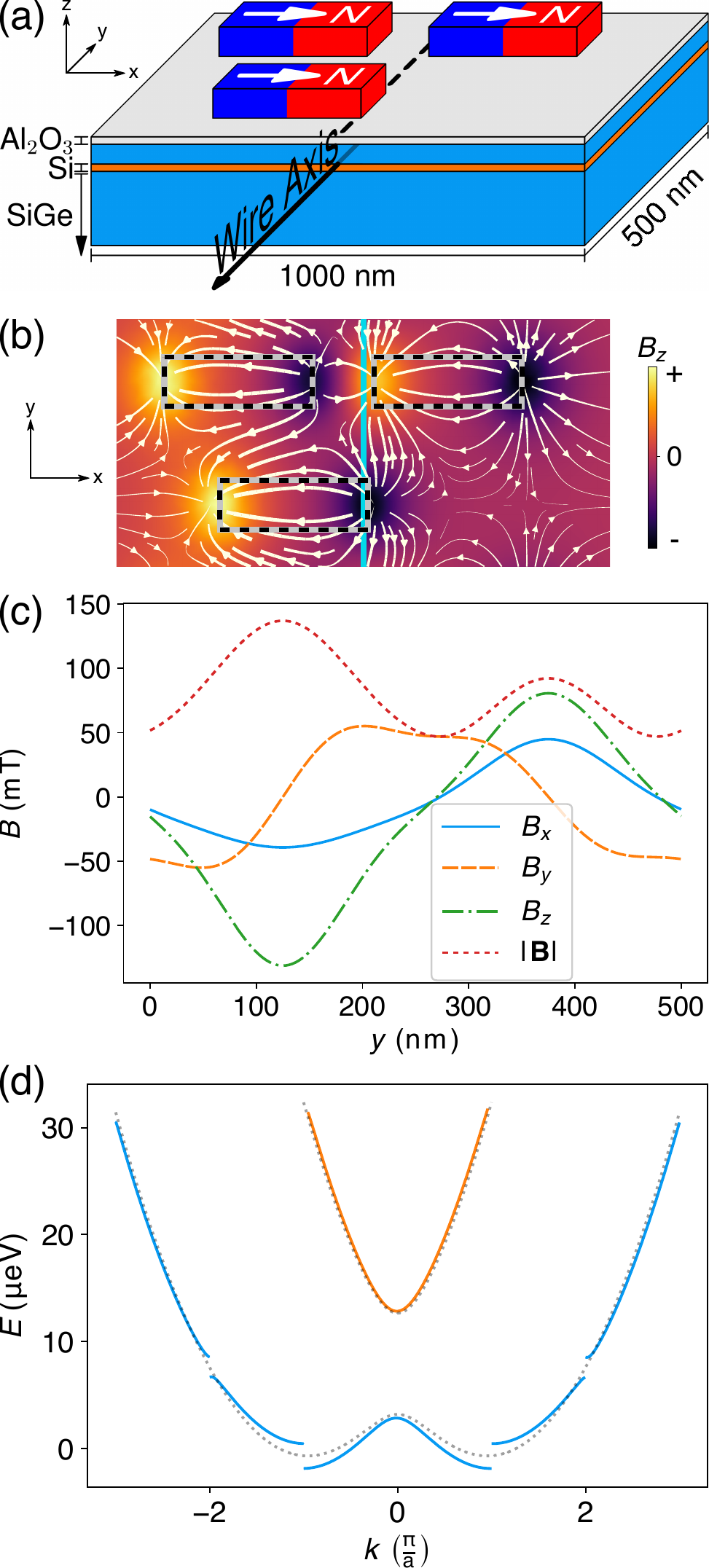}
 \caption{(a) A unit cell of an example single-polarization device. The material stack is the same as in Fig.~\ref{fig:device}. The axis of the wire is indicated. (b) The magnetic field in the plane of the quantum well. Arrows indicate in-plane field. Colors represent out-of-plane field. The wire axis is in blue. The magnets above the well are outlined in dashed black and grey. (c) The magnetic field along the wire axis with 2.2 T magnets. Despite the asymmetry of the magnet placement, the average value of each component is close to zero, and the components have similar amplitudes. (d) The resulting band structure after the unitary transformation. The dotted, light-grey curve shows the band structure resulting from a perfectly sinusoidal magnetic field [Eq.~(\ref{eq:E_SOC})] with $B$ equal to the average $\left|\vect{B}\left(x\right)\right|$. The design shows a clear spin-orbit gap, which is over 5 times larger than the bad gap.} \label{fig:unidir}
\end{figure}

Kjaergaard et al. proposed a design with all the magnets polarized in the same direction \cite{Flensberg_PRB_12}, which could simplify fabrication. Unlike the other designs we have shown, the single-polarization design from Kjaergaard et al. has the magnets in plane with the 1D channel. A similar design is shown in Fig.~\ref{fig:PRB_device}. While the design does create the necessary magnetic field rotation, it also introduces very large gaps due to the large average magnetic field, and the spin-orbit gap is washed out [Fig.~\ref{fig:PRB_device}(c)].  Superconductivity induced by an s-wave superconductor is expected to be very weak in this case, as the large average magnetic field aligns the spin toward the same direction ($+\hat{\vect{x}}$) for the most part of the wire and prevents singlet pairing.

The main challenge of single-polarization designs is to reduce the average magnetic field to an acceptable level. One way to do this is to place the magnets above the quantum well. In designs where the magnets are in plane with the quantum well, it is difficult to create fields at the wire that are opposite to the polarization direction. For example, the design in Fig.~\ref{fig:PRB_device} has a magnet polarized in $+\hat{\vect{x}}$, and Fig.~\ref{fig:PRB_device}(b) shows that $B_x < 0$ only briefly. In contrast, the magnetic field under a magnet is in the opposite direction to the magnet polarization, so placing the magnets above the electron channel allows us to easily generate magnetic fields opposite the magnet polarization. By carefully placing magnets above the electron channel, we can create designs with a small average magnetic field. The gate would be buried below the magnets (not shown).

While designs with two magnets per unit cell can produce acceptable results, we have found that designs with three magnets per unit cell can do a better job of reducing the average magnetic field while keeping the magnetic field components nearly 90 degrees out of phase. One design is shown in Fig.~\ref{fig:unidir}(a). A pair of magnets is located on either side of the wire axis while a single magnet is located above the wire axis. The logic of this design is that it produces a rotating magnetic field in the following way:

\begin{itemize}
	\item The field between the pair of magnets will have a positive x component, while the field below the single magnet will have a negative x component.
	\item The wire axis runs close to the south pole of one magnet and the north pole of another, which produces z components of the magnetic field in opposite directions.
	\item The field lines run from the north pole of the single magnet to the south pole of the rightmost magnet (and to the rightmost magnet in the next unit cell along the wire axis). That causes an oscillating y component of the magnetic field. The y component of the magnetic field is nearly 90 degrees out of phase from the other two components.
\end{itemize}

\noindent
Each component has an average value close to zero, which results in a small average magnetic field.

The design in Fig.~\ref{fig:unidir} produces a spin-orbit gap of 10 {\textmu}eV (116 mK) and a bad gap of only 1.9 {\textmu}eV.

We believe that single-polarization designs with magnets above the channel are very promising for practical reasons. Since the nanomagnets are placed on top of the SiGe heterostructure, no etching of the SiGe material is required.  This eliminates a difficult fabrication step that requires precise etching depth control and degrades mobility. The main disadvantage of designs like the one shown in Fig.~\ref{fig:unidir}(a) is that there is no room for a gate on top of the stack, and the gate would have to be buried below the magnets. However, making a buried gate may be easier than fabricating magnets with two polarizations.

\subsection{Spin Rotation} \label{sec:spin}

\begin{figure}
 \centering
 \includegraphics[width=0.775\columnwidth]{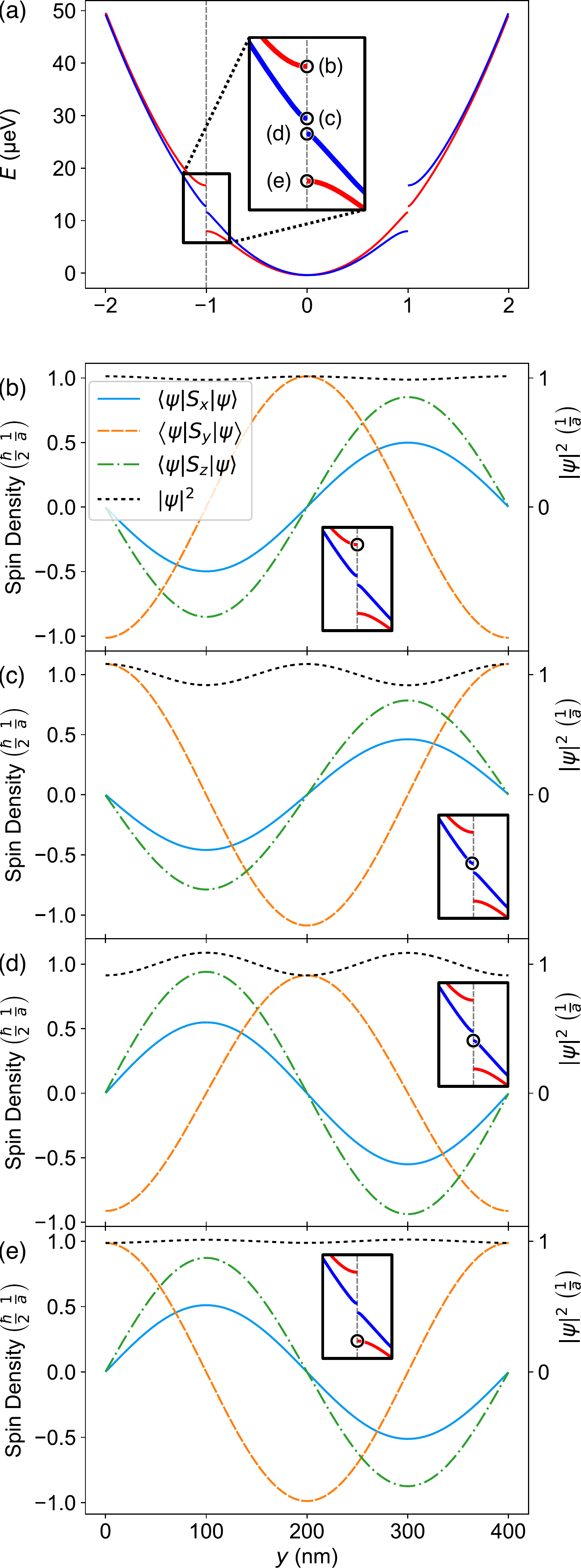}
 \caption{Illustration of spin rotation for the four wave functions with $k = - \sfrac{\pi}{a}$, which is the location of both the good and bad gaps in the untransformed band structure. (a) Band structure for wire in Fig.~\ref{fig:device}. Inset indicates the four modes and the associated subfigures. (b)-(e) show $\left<\psi\left|S_x\right|\psi\right>$ (light blue), $\left<\psi\left|S_y\right|\psi\right>$ (orange), $\left<\psi\left|S_z\right|\psi\right>$ (green), and $\left|\psi\right|^2$ (black, right axis) for the wave functions corresponding to the circled locations on the band structure. ($S_{x,y,z}$ are the spin operators.) In the lowest energy mode (e), the spin expectation value is roughly antiparallel to the magnetic field [Fig.~\ref{fig:device}(c)]. In the highest energy mode (b), the spin expectation value is roughly parallel to the magnetic field. In the two middle-energy modes (c)-(d), the magnetic field and spin expectation values rotate in opposite directions, so $0 \approx \int_0^a \vect{B} \cdot \vect{S} \; dy$.} \label{fig:spin}
\end{figure}

As an electron moves through the structures described here, its spin rotates. The spin rotation is momentum- and band-dependent, as expected for a system with spin-orbit coupling. As an example, in Fig.~\ref{fig:spin} we illustrate the spin rotation in the states nearest the good and bad gaps at $k = -\sfrac{\pi}{a}$ for the device shown in Fig.~\ref{fig:device}. The lowest energy state [Fig.~\ref{fig:spin}(e)] has its spin roughly antiparallel to the magnetic field [Fig.~\ref{fig:device}(c)], while the highest energy state [Fig.~\ref{fig:spin}(b)] has its spin roughly parallel to the magnetic field. In both cases, when viewed from the $+\hat{\vect{z}}$ axis, the magnetic field and the spin expectation values rotate clockwise as we move along the $+\hat{\vect{y}}$ axis.. For the two states with intermediate energies [Fig.~\ref{fig:spin}(c) and (d)], the magnetic field and spin expectation values rotate in opposite directions, so $0 \approx \int_0^a \vect{B} \cdot \vect{S} \; dy$. (If that were exactly zero, there would be no bad gap.) In all cases, the spin rotates with the same period as the magnetic field, and the magnitude of the wave function is fairly constant (black curve, right axis). These results are consistent with the simple physical picture that, when rotating spins are used as the basis, only the branches with the same chirality and period as the rotating magnetic field are strongly perturbed in energy.  The rotation and wave function magnitude can be more complicated for other $k$ values.

\section{Conclusion} \label{sec:conclusion}

Our calculations indicate that a nanomagnet-induced artificial spin-orbit gap should be realizable in a Si/SiGe system, as well as in other nanomagnet systems with 1D channels. By modeling several example devices, we have gained useful insight into how the magnets affect band structure:

\begin{itemize}
  \item Unwanted gaps open unless the first Fourier coefficient of the magnetic field has components that are equal in magnitude and are 90 degrees out of phase [Sec.~\ref{sec:gaps} and Fig.~\ref{fig:device}(d)].
  \item A nonzero average magnetic field will open an additional set of gaps, and if the average magnetic field is too large, the spin-orbit gap will be washed out [Sec.~\ref{sec:strengths} and Fig.~\ref{fig:PRB_device}(c)].
  \item An asymmetric magnetic field can shift the k-values where the extrema of the band structure occur [Sec.~\ref{sec:strengths} and Fig.~\ref{fig:different_strengths}(b)].
\end{itemize}

Despite these obstacles, the band structure of these devices approximates the ideal band structure [Eq.~\ref{eq:E_SOC}] reasonably well for a variety of designs [Figs.~\ref{fig:device}(d), \ref{fig:different_strengths}(c), and \ref{fig:unidir}(d)].

The above insights helped us to design a single-polarization device that should produce a measurable spin-orbit gap (Fig.~\ref{fig:unidir}). Such a design should be much easier to fabricate than alternating-polarization designs, and we hope to present an experimental realization of an optimized single-polarization design in the future.

\begin{acknowledgments}
Sandia National Laboratories is a multimission laboratory managed and operated by National Technology and Engineering Solutions of Sandia LLC, a wholly owned subsidiary of Honeywell International Inc. for the U.S. Department of Energy's National Nuclear Security Administration under contract DE-NA0003525. This work was funded by the Laboratory Directed Research and Development Program. This work was performed, in part, at the Center for Integrated Nanotechnologies, a U.S. DOE, Office of Basic Energy Sciences, user facility. Work at Los Alamos National Laboratory was funded by the US DOE, Office of Science, Basic Energy Sciences, Division of Materials Sciences and Engineering.

The authors thank N. T. Jacobson for helpful comments on this article.
\end{acknowledgments}

\appendix*
\section{Unitary Transformation} \label{app:unitary}

Here we review the position-dependent unitary transform used in \cite{Flensberg_PRB_12}, which has the general form of

\begin{equation} \label{eq:U}
  U\left(y\right) = e^{\frac{1}{2} i \vect{\theta}\left(y\right) \cdot \vect{\sigma} },
\end{equation}

\noindent
where $\vect{\theta} = \theta \hat{\vect{\theta}}$, and the choice of $\hat{\vect{\theta}}$ and $\theta$ are discussed below.

In \cite{Flensberg_PRB_12}, Eq.~(\ref{eq:U}) is used to rotate the $x$ axis \footnote{Ref. \cite{Flensberg_PRB_12} actually uses the $z$ axis instead of the $x$ axis, but the results are equivalent.} of the local spin basis to be in the direction of the magnetic field. (See Ch. 14 of \cite{Shankar_book}. The sign convention is subtly different.) Then,

\begin{equation} \label{eq:variable_theta_hat}
  \hat{\vect{\theta}}\left(y\right) = \frac{\vect{B}\left(y\right) \times \hat{\vect{x}}}{\left|\vect{B}\left(y\right) \times \hat{\vect{x}}\right|}
\end{equation}

\noindent
is a unit vector perpendicular to both $\hat{\vect{x}}$ and $\vect{B}$, and $\theta$ is the angle between $\hat{\vect{x}}$ and $\vect{B}$:

\begin{equation} \label{eq:variable_theta}
  \cos{\theta} = \frac{\vect{B} \cdot \hat{\vect{x}}}{\left|\vect{B}\right|}.
\end{equation}

\noindent
The choice of $\hat{\vect{\theta}}$ and $\theta$ in Eqs.~(\ref{eq:variable_theta_hat}) and (\ref{eq:variable_theta}) can be thought of as putting the Hamiltonian [Eq.~(\ref{eq:H})] into a reference frame that rotates with the magnetic field.

The unitary transformation will make the band structure of our system's Hamiltonian [Eq. (\ref{eq:H})] look like the band structure resulting from a system with strong spin-orbit coupling in an external magnetic field if $\theta\left(y\right)$ varies continuously and $\left|\theta\left(0\right) - \theta\left(a\right) \right| = 2\pi$. For that to happen, the $\hat{\vect{x}}$'s in Eqs. (\ref{eq:variable_theta_hat}) and (\ref{eq:variable_theta}) must be replaced by a unit vector that is parallel to $\vect{B}\left(y\right)$ for some $y$.

Now, consider the special case of the perfectly sinusoidal magnetic field in Sec.~\ref{sec:U}. Inserting the $\vect{B}$ in Eq. (\ref{eq:sinusoidal_B}) into Eqs. (\ref{eq:variable_theta_hat}) and (\ref{eq:variable_theta}) results in

\begin{equation} \label{eq:theta_hat}
  \hat{\vect{\theta}} = \hat{\vect{z}}
\end{equation}

\noindent  
and

\begin{equation} \label{eq:thetat}
  \theta = \frac{2 \pi y}{a}.
\end{equation}

\noindent
Combining Eqs. (\ref{eq:theta_hat}) and (\ref{eq:thetat}) with Eq. (\ref{eq:U}) results in the unitary transformation given in Eq. (\ref{eq:sinusoidal_U}). 

As we note in Sec.~\ref{sec:U}, we use the unitary transformation in Eq. (\ref{eq:sinusoidal_U}) even when the magnetic field is not perfectly sinusoidal. This is because the transformation in Eq. (\ref{eq:sinusoidal_U}) is much simpler than the transform given by Eqs. (\ref{eq:variable_theta_hat}) and (\ref{eq:variable_theta}) for a general magnetic field, but both transforms have nearly the same effect on the band structure; for the case of the structure in Fig.~\ref{fig:device}(a), the band structures produced by the two transforms are visually indistinguishable. In some cases, the differences are visible, but the differences are not meaningful because we are simply using the transform as a tool to view the band structure in a more convenient form.


%

\end{document}